\documentclass[12pt]{article}
\usepackage{latexsym}
\usepackage{amsfonts}
\usepackage{amssymb}
\usepackage{feyn}
\usepackage[cp1250]{inputenc}

\title{Note on charge interaction in NQED}
\author{Katarzyna Bolonek\thanks{supported by the grant 1 P03B 125 29 of the Polish Ministry of Science and  by The European Social Fund and Budget of 
State implemented under The Integrated Regional Operational Programme - Project GRRI-D and the grant 690 of the University of Lodz.},\\ 
 Piotr  Kosi\'nski\thanks{supported by the Polish Ministry of Science Grant.} \\
Department of Theoretical Physics II \\
University of {\L}\'od\'z \\
Pomorska 149/153, 90 - 236 {\L}\'od\'z, Poland.}

\date{}

\begin{document}
\maketitle
\begin{abstract}
The interaction of charges in NQED is discussed. It is shown that the relativistic correction have the same form as in the 
commutative case provided the Weyl ordering rule is used.

\end{abstract}

\newpage
It is well known that the relativistic quantum theory differs from its nonrelativistic counterpart in two important respect. First, the notion 
of particle coordinate is of limited value. This is due to the fact that in order to localize the particle one has to apply the external field, 
the stronger the more localized state one wants to obtain. Then, due to the possibility of pair creation, the number of particles in the systems 
becomes undefined. Therefore, one cannot localize the particle beyond certain limit. Only in the case of very heavy particles strict localization 
is possible which allows to define the space-time points as coordinates of infinitely massive particles (this picture breaks down if gravity is 
taken into account).

Secondly, in relativistic theory all interactions propagate with finite velocity. This makes the notion of interaction energy useless unless 
one takes the nonrelativistic limit. Then the expansion in powers of $\frac{v}{c}$, $v$\ being a typical velocity of charged particle, produces 
succesive corrections to charge interaction \cite {b1}.  

The aim of the present note is to show that the similar situation is encountered in space-space $(Q^{0i}=0)$\ noncommutative quantum 
electrodynamics \cite {b2}. We find that by making an appropriate expansion in powers of $\frac{v}{c}$\ one produces the interaction energy of the 
same form as in commutative case provided it is expressed in terms of noncommutative coordinates (and momenta) by means of Weyl ordening.

We start with quantum mechanics on noncommutative configuration space \cite {b3}. The basic commutation rules read
\begin{eqnarray}
[\hat x_i,\hat x_j]=i\Theta _{ij}, \;\;[\hat x_i,\hat p_j]=i\delta _{ij}, \;\; [\hat p_i,\hat p_j]=0  \label{w1}
\end{eqnarray}
As always, in order to deal with definite representation we look for the maximal set of communiting observables. One such a set consists of all 
momentum components $\hat p_i$. However, one can also find a counterpart of coordinate representation. To this end we define new variables 
$\hat X_i,\hat P_i$\ by
\begin{eqnarray}
\hat x_i\equiv \hat X_i-\frac{1}{2}\Theta _{ik}\hat P_k, \;\;\; \hat p_i=\hat P_i;   \label{w2}
\end{eqnarray}
then $\hat X_i,\hat P_i$\ obey standard commutation rules.

Consider now some function $U(\hat {\vec x},\hat {\vec p})$\ defined according to the Weyl correspondence rule
\begin{eqnarray}
U(\hat {\vec x}, \hat {\vec p})=\int d^3 \vec {\alpha }d^3 \vec {\beta }\tilde U
(\vec {\alpha }, \vec {\beta })e^{i \vec {\alpha }\hat  {\vec  {x}}+
i \vec {\beta } \hat {\vec  {p}}}  \label{w3}
\end{eqnarray}
where
\begin{eqnarray}
\tilde U(\vec {\alpha }, \vec {\beta })=\frac{1}{(2\pi )^6} \int d^3\vec {x}d^3\vec {p}  
U(\vec {x},\vec {p})e^{-i\vec {\alpha }\vec {x}-i\vec {\beta }\vec {p}}  \label{w4}
\end{eqnarray}
is the Fourier transform of the function $U(\vec {x}, \vec {p})$\ defined on classical phase space.

Let $\mid \vec {p}\rangle , \mid \vec {p}'\rangle $\  be the eigenvectors  of $\hat {\vec {p}}\equiv \hat {\vec  
{P}}$. We want to calculate the matrix element
\begin{eqnarray}
\langle \vec {p}'\mid U(\hat {\vec {x}}, \hat {\vec {p}})\mid \vec p\rangle =\int d^3\vec {\alpha }d^3
\vec {\beta }\tilde U(\vec {\alpha }, \vec {\beta })\langle \vec {p}'\mid e^
{i\vec {\alpha }\hat {\vec {x}}+i\vec {\beta }\hat {\vec {p}}}\mid \vec {p}\rangle  \label{w5}
\end{eqnarray}
Now, by virtue of eq. (\ref {w2}) we have
\begin{eqnarray}
&& \langle \vec {p}'\mid e^{i\vec {\alpha } \hat {\vec {x}}+i\vec {\beta }\hat {\vec {p}}}
\mid \vec {p}\rangle =\langle \vec {p}'\mid e^{i\vec {\alpha }\hat {\vec {X}}+
i\vec {\beta }\hat {\vec {P}}}e^{-\frac{1}{2}\Theta _{ij}\alpha _i\hat P_j}\mid \vec {p}\rangle= \nonumber \\
&&=e^{-\frac{1}{2}\Theta _{ij}\alpha _ip_j}\langle \vec p'\mid e^{i\vec \alpha \hat {\vec X}+
i\vec \beta \vec P}\mid \vec p\rangle   \label{w6}
\end{eqnarray}
because $\Theta _{ij}\alpha _i\hat P_j$\ commutes with $\vec \alpha \hat {\vec X}+\vec \beta \hat {\vec P}$. \\
The last term on the RHS of eq.(\ref {w6}) can be written as
\begin{eqnarray}
 \langle \vec p'\mid e^{i\vec \alpha \hat {\vec X}+i\vec \beta \hat {\vec P}}\mid \vec p\rangle =e^{\frac{i}{2}\vec \alpha \vec \beta } 
\langle \vec p'\mid e^{i\vec \alpha \hat {\vec X}}e^{i\vec \beta \hat {\vec P}}\mid \vec p\rangle = 
e^{\frac{i}{2}\vec \alpha \vec \beta }e^{i\vec \beta \vec p }\langle \vec p'\mid e^{i\vec \alpha \hat {\vec X}}\mid \vec p\rangle  \label{w7}
\end{eqnarray}
Now, $\hat X_i$\ form a complete commuting set; consequently,
\begin{eqnarray}
&& \langle \vec p'\mid e^{i\vec \alpha \hat {\vec X}}\mid \vec p\rangle =\int d^3\vec Xd^3\vec X'\langle \vec p'\mid \vec X'\rangle \langle \vec X'\mid 
e^{i\vec \alpha \hat {\vec X}}\mid \vec X\rangle \langle\vec X\mid \vec p\rangle=  \nonumber \\
&& =\frac{1}{(2\pi )^3}\int d^3\vec Xd^3\vec X'e^{i(\vec \alpha +\vec p-\vec p')\vec X}\delta ^{(3)}(\vec X-\vec X')=\delta ^{(3)}
(\vec \alpha +\vec p-\vec p')  \label{w8}
\end{eqnarray}
Eqs.$(\ref {w4})\div (\ref {w8})$\ lead to
\begin{eqnarray}
\langle \vec p'\mid U(\hat {\vec x},\hat {\vec p})\mid \vec p\rangle =e^{\frac{i}{2}\Theta _{ij}p_ip'_j}\int d^3\vec \beta 
\tilde U(\vec p'-\vec p,\vec \beta )e^{\frac{i}{2}\vec \beta (\vec p+\vec p')}  \label{w9}
\end{eqnarray}
The main conclusion of the above reasoning is that the matrix element between momentum eigenstates in noncommutative quantum mechanics differs 
from the one in standard quantum mechanics, by universal multiplicative factor $exp(\frac{i}{2}\Theta _{ij}p_ip_j)$. This result readily 
generelizes to the multiparticle case where one gets one factor of the above form for each particle.

Let us now consider the scattering of two electrons in noncommutative QED with space-space noncommutativity $(\Theta ^{i0}=0))$. To the lowest 
order it is described by two graphs depicted on Fig.1. \\
\begin{figure}
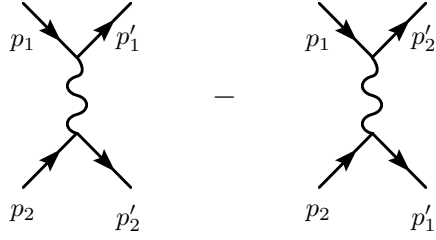

\begin{displaymath}
\large{\Diagram{\vertexlabel^{p_1}  fdA &  fuA \vertexlabel^{p_1'} \\& gv\\ \vertexlabel_{p_2}fuA &  fdA \vertexlabel_{p_2'}\\}\qquad - \qquad\Diagram{\vertexlabel^{p_1}fdA &  fuA \vertexlabel^{p_2'}\\ & gv\\ \vertexlabel_{p_2}fuA &  fdA \vertexlabel_{p_1'}}}
\end{displaymath}
\caption{Electron scattering to the lowest order}
\end{figure}
The corresponding matrix element reads
\begin{eqnarray}
\langle p_1'p_2'\mid S\mid p_1p_2\rangle=\frac{ie^2(\bar u(p_1')\gamma ^{\mu }u(p_1))(\bar u(p'_2)\gamma _{\mu }u(p_2))}{(p_1'-p_1)^2}
e^{\frac{i}{2}p_1\Theta p'_1}e^{\frac{i}{2}p_2\Theta p'_2}-(p_1'\leftrightarrow p_2')   \label{w10}
\end{eqnarray}
Let us first consider the commutative case, $\Theta ^{ij}=0$. 

As it is well known the matrix element cannot be written as a matrix element of some interaction energy operator; this is due to the 
fact that the interaction is not instanteneous. However, if we asume that particle velocities are small we can expand the above matrix element 
in powers of $\frac{v}{c}$. Each term of this expansion can be put in the form of matrix element of some operator describing instanteneous 
interaction. In such a way one obtains succesive corrections to the electron-electron interaction energy \cite {b4}.

To the order $\frac{v^2}{c^2}$\ the interaction energy $(\Theta =0)$\ reads \cite {b4}
\begin{eqnarray}
 && U(r)=  
 \frac{\alpha }{r}-\frac{\pi \alpha }{m^2c^2}\delta ^{(3)}(\vec r)- \label{w11} \\
&& -\frac{\alpha }{2m^2c^2r^3}\left( (\vec r\times \vec p_1)
\vec s_1-(\vec r\times \vec p_2)\vec s_2+ 
2(\vec r\times \vec p_1)\vec s_2-2(\vec r\times \vec p_2)\vec s_1\right)- \nonumber \\ 
&& -\frac{\alpha }{2m^2c^2}\left(\frac{1}{r}\vec p_1\vec p_2+\frac{1}{r^3}\vec r(\vec r\vec p_1)\vec p_2\right)
 +\frac{\alpha }{m^2c^2}
\left(\frac{\vec s_1\vec s_2}{r^3}-\frac{3(\vec s_1\vec r)(\vec s_2\vec r)}{r^3}-\frac{8\pi }{3}\vec s_1\vec s_2\delta ^{(3)}(\vec r)\right)  \nonumber
\end{eqnarray}
where $\vec r\equiv \vec r_1-\vec r_2$\ and $\alpha $\ is the structure constant. As usual, the most singular terms are dealt with by integrating 
over angles first. Moreover, the above expression is already Weyl ordered (although this is not explicitly seen from eq. (\ref {w11})).

Now, according to eq. ({\ref {w10}) the NQED amplitude differs by two exponential factors corresponding to both electrons. By virtue of eq. (\ref {w9}) 
we conclude that, to the order $\frac{v^2}{c^2}$, the interaction energy between electrons in NQED is given by the same expresion (\ref {w11}) 
but with coordinates replaced by their noncommutative counterparts and defined by Weyl ordering rule. 

In the case of space-time noncommutativity $(\Theta ^{0i}\neq 0)$\ one cannot repeat this reasoning. In fact, in order to localize the interaction in time 
(which is necessary to obtain the interaction energy) one should have made also an expansion in powers of $\Theta $.

\end{document}